\documentclass[twocolumn,showpacs,amsmath,amssymb,aps,10pt]{revtex4-1}

\usepackage{epsfig}
\usepackage{verbatim}
\usepackage{dcolumn}
\usepackage{graphicx}
\usepackage{bm}
\usepackage{epstopdf}

\newcommand{\eg}{\emph{e.g.,} }

\newcommand{\be}{\begin{equation}}
\newcommand{\ee}{\end{equation}}
\newcommand{\bea}{\begin{equation*}}
\newcommand{\eea}{\end{equation*}}
\newcommand{\beqr}{\begin{eqnarray} \nonumber}
\newcommand{\eeqr}{\end{eqnarray}}
\newcommand{\beqrb}{\begin{eqnarray}}
\newcommand{\eeqrb}{\nonumber \end{eqnarray}}
\newcommand{\fin}{\mbox{ .}}
\newcommand{\coma}{\mbox{ ,}}

\newcommand{\const}{\mbox{const.}}

\newcommand{\grad}{\bm{\nabla}}
\newcommand{\vect}[1]{\bm{#1}}
\newcommand{\unit}[1]{\bm{\hat{#1}}}
\newcommand{\pr}{\partial}

\newcommand{\myNi}{\emph{(i)}\,}
\newcommand{\myNii}{\emph{(ii)}\,}
\newcommand{\myNiii}{\emph{(iii)}\,}
\newcommand{\myNiv}{\emph{(iv)}\,}

\newcommand\mnras{{MNRAS}} 
\newcommand\aap{{A\&A}}    

\newcommand{\jgr}{Journal of Geophysical Research}
\newcommand{\planss}{Planetary and Space Science}

\def\myfig#1{./#1}

\newcommand{\myS}{{S}}
\newcommand{\myW}{{W}}
\newcommand{\overbar}[1]{\mkern 1.5mu\overline{\mkern-1.5mu#1\mkern-1.5mu}\mkern 1.5mu}
\newcommand{\mybar}[1]{\overbar{#1}}

\newcommand{\overtilde}[1]{\mkern 1.5mu\widetilde{\mkern-1.5mu#1\mkern-1.5mu}\mkern 1.5mu}
\newcommand{\mytilde}[1]{\tilde{#1}}
\newcommand{\myttilde}[1]{\overtilde{#1}}
\newcommand{\mystag}[1]{{#1}_0}
\newcommand{\myGamma}{\gamma}   
\newcommand{\mycs}{c}           
\newcommand{\mycsStag}{\bar{c}} 
\newcommand{\mySO}{\Delta}      
\newcommand{\myg}{g}            

\begin{document}

\title{Compressible flow in front of an axisymmetric blunt object}

\author{Uri Keshet}

\email{ukeshet@bgu.ac.il}

\author{Yossi Naor}


\affiliation{Physics Department, Ben-Gurion University of the Negev, POB 653, Be'er-Sheva 84105, Israel}

\date{\today}

\begin{abstract}
Compressible flows around blunt objects have diverse applications, but present analytic treatments are inaccurate and limited to narrow parameter regimes.
We show that the flow in front of an axisymmetric body is accurately derived analytically using a low order expansion of the perpendicular gradients in terms of the parallel velocity.
This reproduces both subsonic and supersonic flows measured and simulated for a sphere, including the transonic regime and the bow shock properties.
\end{abstract}

\maketitle

Compressible flows around blunt objects play an important role in diverse fields, ranging from fluid mechanics \cite[\eg][]{LandauLifshitz59_FluidMechanics, ParkEtAl2006,MackSchmid2011,TuttyEtAl2013,GrandemangeEtAl2013,GrandemangeEtAl2014}, space physics \cite{SpreiterAlksne70, BaranovLebedev88, SpreiteStaharar95, CairnsGrabbe94, PetrinecRussell97, Petrinec02, Zhangetal04}, and astrophysics \cite{LeaYoung76, ShavivSalpeter82, CantoRaga98, Vikhlininetal01, Lyutikov06, SchulreichBreitschwerdt11}, to computational physics and applied mathematics \cite{Hejranfaretal09, Wilson13, GollanJacobs13, Marroneetal13}, aeronautical and civil engineering \cite{NAKANISHIKAMEMOTO93, Baker10, Aulchenkoetal12}, and aerodynamics \cite{AsanalievEtAl88,LiouTakayama05,PilyuginKhlebnikov06,Volkov09}.
Yet, even for the simple case of an inviscid flow around a sphere, the problem has resisted a general or accurate analytic treatment, due to its nonlinear nature.

For example, in space physics and astrophysics, the interaction of an ambient medium with much denser, approximately solid, bodies such as comets \citep[\eg][]{BaranovLebedev88}, planets \citep{SpreiterAlksne70,CairnsGrabbe94,PetrinecRussell97}, binary companions \citep{CantoRaga98}, galaxies \citep{ShavivSalpeter82,SchulreichBreitschwerdt11}, or large scale clumps and bubbles \citep{LeaYoung76,Vikhlininetal01,Lyutikov06}, is important for modeling these systems and understanding their observational signature. This is particularly true for the shocks formed in supersonic flows, due to their rich nonthermal effects \citep[\eg][]{,SpreiteStaharar95,Vikhlininetal01, Petrinec02,Zhangetal04}.

Such research typically involves an idealized, inviscid flow around a simple blunt object, often approximated as axisymmetric or even spherical.
The flow is sometimes computed numerically, but some approximate analytic description is usually employed in order to gain a deeper, more general understanding of the system.
Consequently, this fundamental problem of fluid mechanics has received substantial attention.
The small Mach number $M$ regime was studied as an asymptotic series about $M=0$ \cite{LordRayliegh1916, tamada39, kaplan40,StangebyAllen71,Allen13}, and solved in the incompressible potential flow limit.
Some hodograph plane results and series approximations were found in the transonic and supersonic cases \cite{Guderley_TransonicFlow, Hida1955asymptotic, LiepmannRoshko57}.
In particular, approximations for the standoff distance of the bow shock \cite{Guy74, Moeckel49, CoronaRomero13, Lighthill57, Hida53, HayesProbstein66} partly agree with experiments \cite{SpreiteStaharar95, SchwartzEckerman56, Veriginetal99, FarrisRussell94} and numerical computations \cite{IgraFalcovitz10, ChapmanCairns03}.

However, these analytic results are typically based on ad-hoc, unjustified assumptions, such as negligible compressible effects, a predetermined shock geometry \cite{Lighthill57, Guy74}, or an incompressible \cite{Hida53} or irrotational \cite{kawamura1950mem} flow downstream of the shock.
Other approaches use slowly converging, or impractically complicated, expansion series \cite{LordRayliegh1916, Hida1955asymptotic, vanDyke58model, vanDyke75Book}.
In all cases, the results are inaccurate or limited to a narrow parameter regime.
A generic yet accurate analytic approach is needed.

We adopt the conventional assumptions of \myNi an ideal, polytropic gas with an adiabatic index $\myGamma$; \myNii negligible viscosity and heat conduction (ideal fluid); \myNiii a steady, laminar, non-relativistic flow; and \myNiv negligible electromagnetic fields.
Typically, these assumptions hold in front of the object, but break down behind it and in its close vicinity.
We thus analyze the flow ahead of the object.
While spatial series expansions and hodograph plane analyses, when employed separately, are of limited use \cite[for reviews, see][]{vanDyke58model,vanDyke75Book}, we find their combination to give good results over the full parameter range.
In particular, we expand the axial flow in terms of the parallel velocity, rather than of distance.
For simplicity, we begin with a sphere, and later and generalize for other objects.

\emph{Flow equations.---}
The flow is governed by the stationary continuity, Euler, and energy equations,
\begin{equation} \label{eq:FlowEquations}
\grad \cdot (\rho\vect{v})=0 \, ; \,\,\,\, (\vect{v}\cdot \grad)\vect{v}=-\frac{\grad P}{\rho} \, ; \,\,\, \,
\vect{v}\cdot \grad \left( \frac{P}{\rho^\myGamma} \right)=0\, ,
\end{equation}
where $\vect{v}$, $P$ and $\rho$ are the velocity, pressure and density.
At a shock, downstream (subscript $d$) and upstream ($u$) quantities are related by the shock adiabat,
\begin{equation} \label{eq:JumpConditions}
\!\frac{\rho_d}{\rho_u} = \frac{v_u}{v_d} = \frac{(\myGamma+1) M_u^2}{(\myGamma-1)M_u^2+2 } \, ; \,\,\, \,
\frac{P_d}{P_u} = \frac{2\myGamma M_u^2+1-\myGamma}{\myGamma+1},
\end{equation}
with $M\equiv v/\mycs$, and $\mycs=(\gamma P/\rho)^{1/2}$ the sound speed \cite[][]{LandauLifshitz59_FluidMechanics}.

Along streamlines, Bernoulli's equation implies that
\begin{equation} \label{eq:Bernoulli}
w + v^2/2  = \mybar{w} = \const \coma
\end{equation}
where $w=\myGamma P/[(\myGamma-1)\rho]$ is the enthalpy, and a bar denotes (henceforth) a putative stagnation ($v=0$) point.
The far incident flow is assumed uniform and unidirectional, so $\mybar{w}$ is the same constant for all streamlines.  Equation~(\ref{eq:Bernoulli}) remains valid across shocks, as $w+v^2/2$ is the ratio between the normal fluxes of energy and of mass, each conserved across a shock. 

\begin{figure*}[ht!]
\centerline{\hspace{3.5cm}\epsfxsize=23cm \epsfbox{\myfig{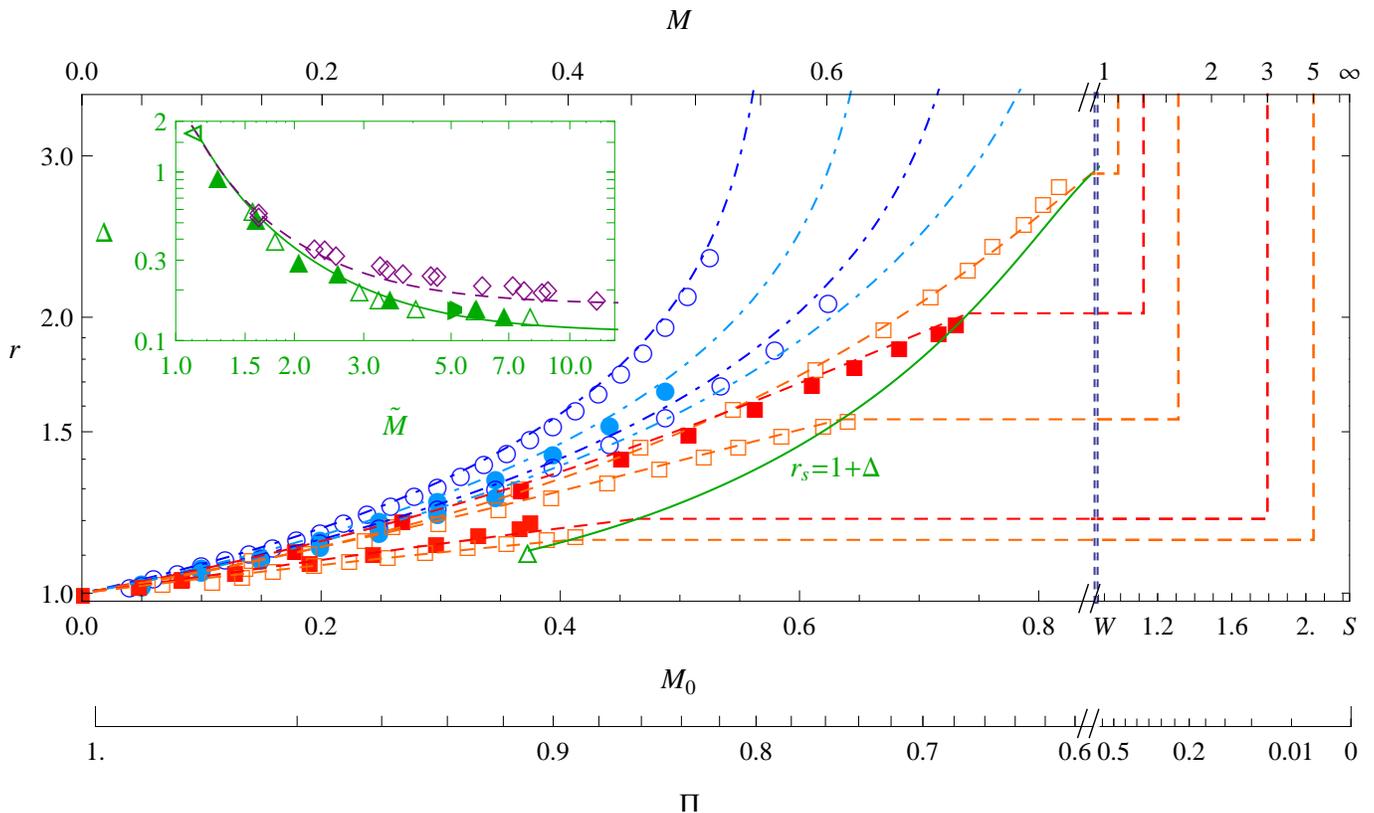}}}
\caption{
Normalized radial profiles of velocity $M_0$ and pressure $\Pi$ (abscissa) in front of a unit ($r=1$) sphere, for $\myGamma=7/5$, according to numerical simulations (symbols) and our approximation (curves), in both subsonic (bluish circles and dot-dashed curves) and supersonic (reddish squares and dashed curves) regimes.
Numerical data shown (alternating shading to guide the eye) for $\myttilde{M}=0.6$, $0.7$, $0.8$, $0.95$ \cite[Ref.][]{Karanjkar08}, $1.1$, $1.3$, $1.62$ \cite{Krause75, Heberle_etal50}, $3$ \cite{BonoAwruch08}, and $5$ \cite{Krause75, SedneyKahl61}.
The shock standoff distance (solid green) with its  $\myttilde{M}\to\infty$ limit (triangle) are also shown.
The right side of the figure extends it (on a different scale, to show the full $M$ range) to the supersonic, $M>1$ part of the flow, upstream of shocks; horizontal jumps represent the shock adiabat Eq.~(\ref{eq:JumpConditions}).
\emph{Inset}: standoff distance measured experimentally (symbols) and using the approximation (curves), for $\myGamma=7/5$ (solid curve and triangle; Refs.~\cite{Heberle_etal50, VanDyke&Milton58, SedneyKahl61, Krause75}; $\beta=0.48$) and $\myGamma=5/3$ (dashed curve and diamonds; Ref.~\cite{SchwartzEckerman56}; $\beta=0.52$).
\label{Fig:AllFlows}
}
\end{figure*}

Bernoulli's Eq.~(\ref{eq:Bernoulli}) relates the local Mach number,
\begin{equation} \label{eq:M0andPi}
M 
= \left(\mystag{M}^{-2}-\myS^{-2}\right)^{-\frac{1}{2}} = ( \Pi^{-\frac{\myGamma-1}{\myGamma}}-1 )^{\frac{1}{2}} \myS \coma
\end{equation}
to the Mach number with respect to stagnation sound, $\mystag{M} \equiv v/\mycsStag$, and to the normalized pressure, $\Pi\equiv P/\mybar{P}$.
We define $\myS^2\equiv 2/(\myGamma-1)$ and $\myW^2\equiv 2/(\myGamma+1)$ as the strong and weak shock limits of $\mystag{M}^2$, so the subsonic (supersonic) regime becomes $0<\mystag{M}<\myW$ ($\myW<\mystag{M}<\myS$).
Figure~\ref{Fig:AllFlows} illustrates these definitions, and shows the shock adiabat (as horizontal $r(M_0)$ jumps; see Eq.~\ref{eq:JumpConditions}) for $\gamma=7/5$.

Consider the flow ahead of a sphere along the symmetry axis, $\theta=0$ in spherical coordinates $\{r,\theta,\phi\}$.
Here, the flow monotonically slows with decreasing $r$, down to $v=0$ at stagnation which we normalize as $\mybar{\vect{r}}=\{1,0,0\}$.
Symmetry implies that along the axis $\vect{v}=-u(r)\unit{r}$, where $u>0$. Here, Eqs.~(\ref{eq:FlowEquations}) become
\begin{equation}
\label{eq:AxisEquations}
\frac{\pr\ln (\rho u)}{\pr\ln r^2}=\frac{q-u}{u}  \, ; \quad
\pr_r P=-\rho u\pr_r u \, ; \quad \pr_\theta P=0 \coma
\end{equation}
along with Bernoulli's Eq.~(\ref{eq:Bernoulli}), where we defined $q\equiv \pr_\theta v_\theta$ as a measure of the perpendicular velocity. Hence,
\begin{align} \label{eq:uODE}
\pr_r u & = \frac{2}{r}(q-u) \frac{1-\mystag{M}^2/\myS^2}{1-\mystag{M}^2/\myW^2} \fin
\end{align}

Our analysis relies on $u(r)$ being a monotonic function.
This allows us to write $q=q(u)$ as a function of $u$ and not of $r$.
Integrating Eq.~(\ref{eq:uODE}) thus yields
\begin{equation} \label{eq:uSolution}
2\ln r = \int_0^{u(r)} \frac{1-\mystag{M}(u')^2/\myW^2}{1-\mystag{M}(u')^2/\myS^2}
\,\, \frac{du'}{q(u')-u'} \, ,
\end{equation}
so given $q(u)$, the near-axis flow is directly determined.

Unlike $u(r)$, or other expansion parameters used previously, the $q(u)$ profile for typical bodies varies little, and nowhere vanishes.
It is well approximated by a few terms in a power expansion of the form
\begin{equation} \label{eq:qExpansion}
q(u) = q_0 + q_1 (u-U) + q_2(u-U)^2 + \ldots \coma
\end{equation}
where $U$ is a reference velocity,
so the integral in Eq.~(\ref{eq:uSolution}) can be analytically carried out to any order.
Moreover, we next show that the boundary conditions tightly fix $q(u)$,
giving a good approximation for the near axial flow.

First expand $q\simeq \mybar{q}$ near stagnation, with $U=\mybar{u}=0$.
An initially homogeneous subsonic or mildly supersonic \cite{LandauLifshitz59_FluidMechanics} flow remains irrotational, $\nabla\times \vect{v}=0$, in which case the lowest-order constraint is
\begin{equation} \label{eq:ConstPotentialFlow}
\mybar{q}_1 = -1/2 \coma
\end{equation}
whereas for a supersonic, rotational flow, it becomes
\begin{equation} \label{eq:ConstGeneralFlow}
3\mycsStag^2\mybar{q}_3+7\mycsStag\mybar{q}_2 = 2\mybar{q}_1 + 6\frac{\mybar{q}_0}{\mycsStag} + \mybar{q}_1\left(\frac{\mybar{q}_0}{\mycsStag}\right)^2 + \left(\frac{\mybar{q}_0}{\mycsStag}\right)^3 \coma
\end{equation}
as seen by expanding Eqs.~(\ref{eq:FlowEquations}) to order $\theta^2(r-1)^3$.
The generalization for non-spherical objects is discussed below.
Next, we estimate $q$ far from the body.

\emph{Subsonic flow.---}
To derive the axial flow out to $r\to\infty$, we use the incident flow (labeled by a tilde, henceforth) boundary condition $\mytilde{\vect{v}}=\mytilde{u}\{-\cos\theta,\sin\theta,0\}$ to expand $\myttilde{q}$ with $U=\mytilde{u}$, such that
\begin{equation} \label{eq:ConstSubsonicQ0}
\myttilde{q}_0= (\pr_\theta \mytilde{v}_\theta)_{\theta=0} =\mytilde{u} \fin
\end{equation}
Additional terms can be derived using $\mytilde{M}\ll 1$ or $r\gg 1$ expansions appropriate for the relevant object.
Here, it will suffice to consider the leading, $(u-\mytilde{u})\propto r^{-\alpha}$ behavior at large radii, such that Eq.~(\ref{eq:uODE}) yields
\begin{equation} \label{eq:ConstSubsonicQ1}
\myttilde{q}_1 = 1 - \frac{\alpha}{2}\, \frac{1-\myttilde{M}_0^2/\myW^2}{1-\myttilde{M}_0^2/\myS^2} \coma
\end{equation}

In the incompressible limit, $\alpha=3$ \cite[\eg][]{LandauLifshitz59_FluidMechanics}, also valid for general forward-backward symmetric objects in any potential flow.
To see the latter, expand the potential $\Phi$, defined by $\vect{v}=\tilde{u}\nabla\Phi$, as a power series in $r$.
The $r\to\infty$ boundary conditions and regularity across $\theta=0$ yield
\begin{align}
\Phi = -r\cos\theta+\frac{\varphi_1}{r \Theta}+\frac{\varphi_2\cos\theta}{r^2 \Theta^3} + \ldots \coma
\end{align}
where $\Theta\equiv [1-M^2(\myS^{-2}+\sin^2\theta)]^{1/2}$.
The constants $\varphi_k$ are determined by the boundary conditions on the specific body.
Symmetry under forward-backward inversion, $\Phi\to-\Phi$ if $\theta\to \pi-\theta$, requires that $\varphi_1=0$.
In general $\varphi_2\neq 0$, implying that indeed $\alpha=3$.
Such behavior is demonstrated for an arbitrary compressible flow around a sphere by the Janzen-Rayleigh series \cite{tamada39,kaplan40}.

Finally, the $\myttilde{q}$ expansion at $r\to\infty$ is matched to the $\mybar{q}$ expansion at stagnation for a potential flow. In the limit of an incompressible flow around a sphere, $q(u)=\tilde{u}-(u-\tilde{u})/2+0(u-\tilde{u})^2=3\tilde{u}/2-u/2$, obtained from Eqs.~(\ref{eq:ConstPotentialFlow}), (\ref{eq:ConstSubsonicQ0}) and (\ref{eq:ConstSubsonicQ1}), is indeed the exact solution.

This procedure reasonably approximates arbitrary compressible, subsonic flows. Better results are obtained by noting that the constraint (\ref{eq:ConstPotentialFlow}) holds beyond stagnation, as long as $\pr_{\theta\theta}v_r$ is negligible, requiring that $\mybar{q}_2\simeq 0$. Combining this with constraints~(\ref{eq:ConstPotentialFlow}), (\ref{eq:ConstSubsonicQ0}) and (\ref{eq:ConstSubsonicQ1}) yields an accurate, third order approximation, shown in Fig.~\ref{Fig:AllFlows}.

\emph{Supersonic flow.---}
Here, a detached bow shock forms in front of the object, at the so-called standoff distance $\mySO$ from its nose.
The transition between subsonic and supersonic regimes is continuous, so $\mySO\to\infty$ as $\myttilde{M}\to 1$, or equivalently as $\myttilde{M}_0\to \myW$.
The unperturbed upstream flow and the shock are shown on the right side of Fig.~\ref{Fig:AllFlows}.

Consider the axial flow between the shock and stagnation.
The $q(u)$ profile is strongly constrained if the normalized shock curvature $\xi^{-1}\equiv (R/r_s)_{\theta=0}$ is known.
Here, $r_s$ is the shock radius, such that $r_s(\theta=0)=1+\mySO$, and $R=-r_s^2/r_s''(\theta)$  is its local radius of curvature.

Expanding the flow Eqs.~(\ref{eq:FlowEquations}) using Eqs.~(\ref{eq:JumpConditions}) as boundary conditions, yields the $q^{(d)}$ expansion coefficients around $U=u_d$, just downstream of the shock,
\begin{equation} \label{eq:q0d}
q_{0}^{(d)} = \myg^{-1}\tilde{u}\left(1+\myg \xi-\xi\right) \, ;
\end{equation}
\begin{equation} \label{eq:q1d}
q_{1}^{(d)} = \frac{3+(\myg-3)\xi}{2}-\frac{1+(3\myg-1)\xi}{1+\myg+(\myg-1)\myGamma} \, ;
\end{equation}
and
\begin{flalign} \label{eq:q2d}
q_{2}^{(d)}  = & \frac{\myg\xi \myW^2}{8\myttilde{u} \left(\myg+\myW^2-1\right)^2} \Big[\frac{ \myg^2-4 \myg+3}{\xi\myW^2} -\frac{ 2 (3\myg+1)}{\xi}  \\
 &+  2 \left(\myg^2+4 \myg+1\right) -\frac{(\myg-1)^2 (\myg+3)}{\myW^2}+\frac{8 \myg^2 \myW^2}{\myg-1} \Big] \coma  \nonumber
\end{flalign}
where $\myg\equiv (\myttilde{M}_0/\myW)^2\geq 1$ is the axial compression ratio.

These coefficients depend on the shock profile only through $\xi$; higher order terms are sensitive to its deviation from a sphere of radius $R$.
In the weak shock limit $\myg\to 1$, so $\xi$ must vanish for $q_{2}^{(d)}$ to remain finite.
Here, $R$ diverges faster than $\Delta$, and $q_{1}^{(d)}\to 1-2\xi$ asymptotes to unity, consistent with a smooth transition to the subsonic regime.
Moreover, if we require that $q_2^{(d)}\to \tilde{q}_2$ in the  $\myttilde{M}_0\to\myW$ limit, then $\xi\to (4+\myGamma)(-1+\myttilde{M}_0/\myW)$, so $R/r_s$ diverges as $(\myttilde{M}_0-\myW)^{-1}$, consistent with \cite{Hida53, Hida1955asymptotic}.

Equations~(\ref{eq:q0d})--(\ref{eq:q2d}) yield a good, second order approximation to the flow, as shown in Fig.~\ref{Fig:AllFlows}, once $\xi$ or any of the $q^{(d)}$ coefficients are determined.
This can be done using the stagnation boundary conditions, such as Eq.~(\ref{eq:ConstGeneralFlow}), but is laborious and body-specific due to the high order involved.
A simpler approach is to estimate $\xi(M)$, which is well approximated by a single-parameter fit over the entire $M$ range.

In the strong shock, $\myttilde{M}_0\to\myS$ limit, the curvature of the shock approaches that of the object \cite{Guy74}; $\xi\to 1$ in the case of the sphere. This, and direct measurements of $\xi$ \cite{Heberle_etal50}, motivate a power-law approximation of the form
\begin{equation} \label{eq:xiFit}
\xi \simeq \left[(\myttilde{M}_0-\myW)/(\myS-\myW)\right]^\beta \coma
\end{equation}
where a small, $\beta<1$ power-law index is needed to reproduce the steep behavior at $\myttilde{M}_0\to\myW$.
Figure \ref{Fig:AllFlows} shows that Eqs.~(\ref{eq:q0d})--(\ref{eq:xiFit}) nicely fit the measured flow throughout the supersonic regime, with $\beta\simeq 1/2$.

The figure inset shows that a single $\beta\simeq 1/2$ power-law reproduces the measured standoff distance throughout the entire Mach range, for two equations of state. Here, $\Delta$ is sensitive to the precise value of $\beta$ only in the $M\simeq 1$ limit; best results are obtained with $\beta=0.48$ ($\beta=0.52$) for $\gamma=7/5$ ($\gamma=5/3$).

\emph{Discussion.---}
The compressible, inviscid flow in front of a blunt object is approximated analytically, using a hodograph-like, $\vect{v}\simeq (-u, q(u)\theta,0)$ transformation.
The velocity (Eq.~\ref{eq:uSolution}) and pressure (Eq.~\ref{eq:M0andPi}) profiles are derived by expanding $q$ as a (rapidly converging) power series in $u$ (Eq.~\ref{eq:qExpansion}), using the constraints imposed by the object (Eqs.~\ref{eq:ConstPotentialFlow} or \ref{eq:ConstGeneralFlow} for a sphere) and by the far upstream subsonic (Eqs.~\ref{eq:ConstSubsonicQ0}--\ref{eq:ConstSubsonicQ1}) or supersonic (Eqs.~\ref{eq:q0d}--\ref{eq:q2d}) flow.

Figure \ref{Fig:AllFlows} shows that a low order $q(u)$ expansion suffices to recover the measured flow in front of a sphere.
The supersonic results also reproduce the measured standoff distance (solid curve and figure inset) of the shock, and constrain its curvature (Eq.~\ref{eq:xiFit}).
Higher-order constraints can be used to improve the approximation further; here we used only the lowest-order constraint at stagnation, and only in the subsonic case.

The axial approximation directly constrains the flow beyond the axis and along the body, as it determines the perpendicular derivatives. For example, one can use it to estimate
$\pr_{\theta\theta}P = -\rho_0 [q^2-u\pr_r(rq)](1-\mystag{M}^2/S^2)^{1/(\Gamma-1)}$, found by expanding Eqs.~(\ref{eq:FlowEquations}) to $\theta^2$ order.
Extrapolation beyond the axis is simpler in the potential flow regime, where, in particular, $\pr_{\theta\theta}v_r=\pr_r(rq)$.

The axial analysis is generalized for any blunt, axisymmetric object, by modifying the $q$ boundary conditions.
For a body with radius of curvature $R_b>0$ at a stagnation radius $r_b$, take $\{z\equiv r\cos\theta=R_b-r_b, \varrho\equiv r\sin\theta=0\}$ as the origin, and rescale lengths by $R_b$.
This maps the stagnation region of the body onto that of the unit sphere, so Eqs.~(\ref{eq:Bernoulli}--\ref{eq:ConstPotentialFlow}, \ref{eq:ConstSubsonicQ0}--\ref{eq:q2d}) remain valid.
The subsonic analysis is unchanged; for an asymmetric body, $\alpha$ may need to be altered, \eg using the Janzen-Rayleigh series.
The supersonic high order Eq.~(\ref{eq:ConstGeneralFlow}) should be adapted for the specific body, or replaced by the fit Eq.~(\ref{eq:xiFit}).
For highly aspherical bodies, the latter may require a modification, fine-tuning $\beta$ to give good results in the weak shock limit, and examining the $\xi\to1$ strong shock limit.

It may be possible to generalize our hodograph-like analysis even for a non-axisymmetric object, using the stagnant streamline instead of the symmetry axis, as long as the corresponding $u$ profile remains monotonic.

\acknowledgements

We thank Ephim Golbraikh and Yuri Lyubarsky for helpful advise.
This research has received funding from the European Union Seventh Framework Programme (FP7/2007-2013) under grant agreement n\textordmasculine ~293975, from IAEC-UPBC joint research foundation grant 257, and from an ISF-UGC grant.


%

\end{document}